# Coexisting Euler and Stiefel-Whitney Topological Phases in Elastic Metamaterials

Jijie Tang[1, #], Adrien Bouhon[2, 4, #], Yue Shen[3], Kailun Wang[3], Junrong Feng[1], Feng Li[1, ‡], Di Zhou[1, †], Robert-Jan Slager[4, 5 $], Ying Wu[3, *]

[1]*Key Lab of Advanced Optoelectronic Quantum Architecture and Measurement (MOE), School of Physics, Beijing Institute of Technology, Beijing 100081, China*

[2]*NORDITA, Stockholm University and KTH Royal Institute of Technology, Stockholm SE-106 91, Sweden*

[3]*School of Physics, Nanjing University of Science and Technology, Nanjing 210094, China*

[4]*Department of Physics and Astronomy, University of Manchester, Oxford Road, Manchester M13 9PL, United Kingdom*

[5]*TCM Group, Cavendish Laboratory, University of Cambridge, Cambridge, UK.*

[#]These authors contributed equally to this work.

[‡]Email: phlifeng@bit.edu.cn

[†]Email: dizhou@bit.edu.cn

[$]Email: robert-jan.slager@manchester.ac.uk

[*]Email: yingwu@njust.edu.cn

**The study of topological band theory in classical structures has led to the development of novel topological metamaterials with intriguing properties. While single-gap topologies are well understood, recent novel multi-gap phases have garnished increasing interest. These novel phases are characterized by invariants, such as the Euler and second Stiefel-Whitney classes, which involve Bloch eigen-subspaces of multiple bands and can change by braiding in momentum space non-Abelian charged band degeneracies belonging to adjacent energy gaps. Here, we theoretically predict and experimentally demonstrate that two of such topological phases can coexist within a single system using vectorial elastic waves. The inherent coupling between different polarization modes enables non-Abelian braiding of nodal points of multiple energy band gaps and results in coexisting Euler and Stiefel-Whitney topological insulator phases. We furthermore unveil the central role played by the topologically stable Goldstone modes' degeneracy. Our findings represent the first realization of hybrid phases in vectorial fields exhibiting topologically nontrivial Goldstone modes, paving the way for bifunctional applications that leverage the coexistence of topological edge and corner states.**

*Introduction.* Topological insulators (TIs) [1-3] are materials that exhibit insulating bulk states but conduct electricity on surfaces through edge states protected by topological invariants, making them particularly intriguing for fundamental physics research. The robust edge states are resistant to scattering and defects, which offers significant potential for applications in lossless wave propagation and control. Over the past decade, there has been substantial progress in understanding TI and Weyl phases [4-12]. More recently, multi-gap topologies have emerged [13-17], where multi-band subspaces (i.e., groups of bands of a given eigen-energy window) can acquire novel invariants that can be manipulated by braiding band degeneracies of successive energy gaps in the momentum space [15, 16]. Specifically, in systems satisfying the spinless space–time inversion symmetry $I_{ST}$ that squares to identity, the Bloch wave functions can always be given real values [14], and the Euler class arises as an integer topological invariant that classifies the vector bundle of real Bloch wave functions associated to *two-band* subspaces, i.e., groups of bands. A nonzero Euler class $e_1$ indicates that the band crossing points between the two bands, similar to vortices, carry the same non-Abelian topological charges [16-20] and hence are topologically obstructed to annihilate. Akin to the Euler class, the second Stiefel-Whitney class $w_2$ characterizes systems with $I_{ST}$-symmetry beyond groups of two bands and physically corresponds to a $\mathbb{Z}_2$ monopole charge [15, 21, 22]. More precisely, in a crystalline system, any two-band subspace hosting an odd Euler class $e_1$ transforms into a multi-band subspace with a nontrivial second Stiefel-Whitney class $w_2 = e_1 (\mod 2) = 1$ when additional trivial bands are included to the same band subspace. A two-dimensional (2D) insulator with $w_2 = 1$ can be referred to as Stiefel-Whitney TI, which supports higher-order corner states in the presence of additional chiral symmetry [15].

To date, multi-gap topologies have been primarily linked to homotopic structures [13, 14], but are increasingly associated with real physical systems. These topologies have been predicted to exhibit novel signatures in out-of-equilibrium systems [23-25], observed in trapped ion insulators [23], and phonon spectra of real materials [22]. Additionally, they have been connected to electronic spectra and magnetic properties [26-28, 8]. The most significant experimental progress has been made in the context of scalar-wave metamaterials [18-19, 21, 28-31]. In contrast, vectorial elastic wave systems offer great versatility and applicability, enabling control over various wave modes in solids [32, 33] with potential for vibration management [34], multi-frequency tuning [35], and dynamic adjustability [36] in engineering applications. Notably, when designing elastic metamaterials, researchers often avoid modes couplings that obstruct the analysis of topological properties. The full vectorial features of elastic waves remain underexplored. Given these considerations, it is natural to ask whether the concepts of Euler, Stiefel-Whitney, and even coexisting topological phases can be realized within vectorial elastic systems.

Accordingly, we theoretically and experimentally demonstrate a hybrid TI that incorporates both Euler and Stiefel-Whitney classes within a single elastic metamaterial. This approach contrasts with the results in Ref. [37], which achieves a hybrid phase based on the Chern class. The fully coupled in-plane and out-of-plane modes enable non-Abelian braiding of multiple-gap band nodes, creating hybrid Euler and Stiefel-Whitney TI phases. Crucially, we show the resultant 2D Euler and Stiefel-Whitney topologies to be intrinsically associated to the topological stability of the crossing of two Goldstone modes at Γ. The nontrivial topologies of these hybrid TI phases are characterized by an odd Euler number and a nonzero second Stiefel-Whitney number, leading to the coexistence of 1D edge states and 0D corner states. Both the edge and corner states have been verified numerically and experimentally with excellent agreement.

*Theoretical Model.* We establish an idealized elastic model that exhibits both Euler and Stiefel-Whitney topological phases. As illustrated in Fig. 1(a), the model features a kagome lattice with two additional layers of triangular lattices positioned above and below. Each layer contains three masses labeled $A_i$, $B_i$, and $C_i$ for $i = 1, 2, 3$, representing the respective layer. Each mass has six degrees-of-freedoms (DOFs), comprising translational and rotational motions that ultimately result in in-plane and out-of-plane collective modes. The masses are interconnected by 3D Timoshenko beams [38-39], including intra-layer nearest-neighbor (NN), next-nearest-neighbor (NNN), third-nearest-neighbor (TNN), and inter-layer coupling matrices, denoted as $D_1$, $D_2$, $D_3$, and $D_4$, respectively. The NNN couplings on layers 1 and 3 ensure the inversion symmetry $I$ while breaking the basal mirror symmetry $\sigma_h$, resulting in coupled in-plane ($xoy$) and out-of-plane ($z$) modes that give rise to the vectorial nature of the field dynamics in this elastic model.

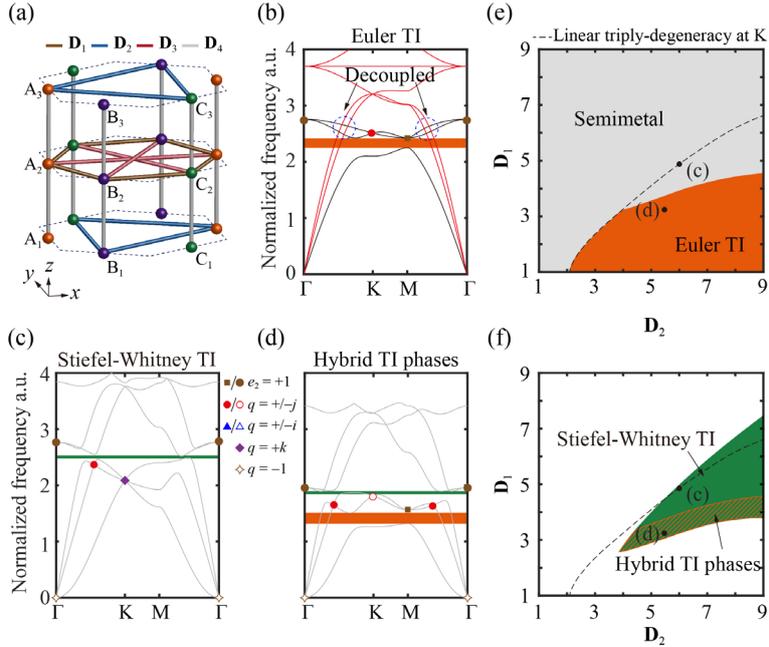

FIG. 1. Mass-beam model. (a) Unit cell of the kagome lattice with colored spheres representing lumped masses connected by 3D beams. (b) Band structure of the mass-spring model. The red and black solid lines represent in-plane and out-of-plane modes, respectively. The quaternion charges of the degenerate points are highlighted by markers. (c) and (d) display the band structures of the idealized mass-beam model with finite bending stiffness, exhibiting the Stiefel-Whitney and hybrid TI phases, respectively. Orange and green regions highlight the Euler and Stiefel-Whitney phases. (e) and (f) show topological phase diagrams as $D_1$ and $D_2$ vary. The black markers correspond to the cases presented in (c) and (d).

Governed by the $9d$ ($d = 6$) DOFs in the unit cell, the model in Fig. 1(a) manifests $9d$ elastic dispersion curves in the momentum space. We focus on the lowest three dispersion curves, which exhibit two-gap degeneracies associated, upon their non-Abelian braiding, with the transitions in both Euler and Stiefel-Whitney topological phases. Under the spinless space–time inversion symmetry $I_{ST}$, the three associated Bloch elastic eigenvectors $\hat{\mathbf{u}}_{1,\mathbf{k}}(\mathbf{x})$, $\hat{\mathbf{u}}_{2,\mathbf{k}}(\mathbf{x})$, $\hat{\mathbf{u}}_{3,\mathbf{k}}(\mathbf{x})$ (i.e. the cell-periodic part of the Bloch wave vectors which dimension is dictated by the mass-beam model together with the Bloch periodicity conditions, see Supplementary Materials) can be represented as strictly real-valued vectors, thereby forming an orthonormal partial frame $\mathbf{R}(\mathbf{x},\mathbf{k}) = \begin{bmatrix} \hat{\mathbf{u}}_{1,\mathbf{k}}(\mathbf{x}) & \hat{\mathbf{u}}_{2,\mathbf{k}}(\mathbf{x}) & \hat{\mathbf{u}}_{3,\mathbf{k}}(\mathbf{x}) \end{bmatrix} \in \mathbb{R}^{54} \times \mathbb{R}^3$. The partial-frame projection $F(\mathbf{k}) = \langle \mathbf{R}(\mathbf{x},\mathbf{k}+\delta\mathbf{k}) | \mathbf{R}(\mathbf{x},\mathbf{k}) \rangle$ (i.e. the $L^2$-inner product that integrates over $\mathbf{x}$, see Supplementary Materials) along a path $\gamma:[\mathbf{k} \leftarrow \mathbf{k}_0]$ then defines the 3D orthonormal frame $W(\mathbf{k}) = \prod_{\mathbf{k} \in \gamma} F(\mathbf{k}) = F(\mathbf{k}) \cdots F(\mathbf{k}_0)$, i.e. an element of $SO(3)$ the group of rotations in the 3D space, to which one associates an element of the $SO(3)$ Lie algebra $I(\mathbf{k}) = \log W(\mathbf{k}) = \int_{\mathbf{k}_0}^{\mathbf{k}} A(\tau) dl$ [19], where $A(\mathbf{k}) = \log F(\mathbf{k}) = \theta_x(\mathbf{k})L_x + \theta_y(\mathbf{k})L_y + \theta_z(\mathbf{k})L_z$ with the 3D rotation generators $L_{x,y,z}$ is analogue to the non-Abelian Berry connection [40]. As the wave vector $\mathbf{k}$ varies, the frame $W(\mathbf{k})$ rotates. In particular, when the dispersion curves intersect at certain wave vectors, the Bloch eigenstates accumulate a geometric phase $\Theta_l = \sqrt{\Theta_x^2 + \Theta_y^2 + \Theta_z^2}$ over a loop $l$ surrounding the nodes [41], obtained from the path integral of the non-Abelian connection $\oint_l A(\mathbf{k}) d\mathbf{k} = \Theta_x L_x + \Theta_y L_y + \Theta_z L_z$ [8]. More generally, the rotating frame $W(\mathbf{k})$ defines an accumulated $SU(2)$ Lie algebra element $\bar{I}(\mathbf{k}) = \int_{\mathbf{k}_0}^{\mathbf{k}} \bar{A}(\tau) dl$ obtained from the $SU(2)$ connection $\bar{A}(\mathbf{k}) = -\frac{i}{2}\left[\theta_x(\mathbf{k})\sigma_x + \theta_y(\mathbf{k})\sigma_y + \theta_z(\mathbf{k})\sigma_z\right]$ using the isomorphism $SU(2) \cong SO(3)$ [16]. This culminates in the definition of an $SU(2)$ frame charge $q = e^{\bar{I}[l]}$ [8,16,20] that corresponds to the faithful monodromic representation of the 1D homotopy class of the Bloch band subspace. In the three-band case the 1D homotopy group is given by the noncommutative quaternion group [8,16,20], $q \in \{\mathbf{1}, \pm i\sigma_x, \pm i\sigma_y, \pm i\sigma_z, -\mathbf{1}\} \leftrightarrow \{1, \pm i, \pm j, \pm k, -1\}$. The non-Abelian frame charges $i$, $j$, $k$ satisfy the fundamental multiplication rules $ij = -ji = k$, $jk = -kj = i$, $ki = -ik = j$, $i^2 = j^2 = k^2 = ijk = -1$, where $q = 1$ represents the trivial frame charge. In particular, the $SU(2)$-quaternion charge $q = -1$ corresponds to the $SO(3)$-geometric angle $\Theta_l = 2\pi$ [8]. Below, we utilize these non-Abelian frame charges to describe the non-Abelian braiding of the band structures, which induces hybrid Euler and Stiefel-Whitney phases in the elastic metamaterial.

We first demonstrate the Euler TI phase using a 3D mass-spring model with fully decoupled in-plane and out-of-plane modes (see Supplementary Materials). The in-plane and out-of-plane stiffnesses are set as $k_{xyNN} = k_{xyNNN} = k_{xyTNN} = 30$ and $k_{zNN} = k_{zNNN} = 25$, $k_{zTNN} = 8$, respectively. The dispersion curves of the three out-of-plane modes are shown by the black curves in Fig. 1(b) and correspond to three-component Bloch eigenvectors $\mathbf{u}_i(\mathbf{k}) = \begin{bmatrix} u_i^x(\mathbf{k}) & u_i^y(\mathbf{k}) & u_i^z(\mathbf{k}) \end{bmatrix}$, $i = 1, 2, 3$. The lowest band maintains a bandgap, whereas the second and third bands exhibit crossing points at certain wavevectors with non-Abelian frame charges. These quaternion charges collectively manifest a topologically nontrivial rotation for the 3D eigenstate frame, which can be described by the non-Abelian Wilson-loop or the topological Euler class, defined as [8,15],

$$e_1 = \frac{1}{2\pi} \int_{BZ} \mathbf{u} \cdot \left( \partial_{k_x} \mathbf{u} \times \partial_{k_y} \mathbf{u} \right) d^2k \qquad (1)$$

where $\mathbf{u}(\mathbf{k})$ is the Bloch wave function for the first band expressed on an orthogonal basis. Numerical calculations yield an odd Euler class, $e_1 = 1$, which also correlates with the meron pattern of the first band [29].

Moreover, the total Euler class of $e_1 = 1$ prevents all nodes with non-Abelian frame charges ($q = +j$ at K point, $e_1 = +1$ at M and Γ points) on the second and third bands from being annihilated together to open a gap between bands 2 and 3. We also note that the lowest two in-plane bands (red curves) intersect the out-of-plane bands around the Γ point. As the in-plane and out-of-plane modes are fully decoupled, such intersections do not introduce an additional band gap.

Having introduced the Euler TI phase in the mass-spring model, we now explore whether the distinct Stiefel-Whitney topological class can emerge in the vectorial mass-beam model when the bending and shearing stiffnesses are considered. The band structure based on the mass-beam model is shown in Fig. 1(c), with $\mathbf{D}_1 = [4.85]$, $\mathbf{D}_2 = [6.0]$, $\mathbf{D}_3 = [4.0]$, and $\mathbf{D}_4 = [7.0]$, where [·] denotes a matrix. A complete bandgap (marked by the green region) emerges between the third and fourth bands. In the mass-spring system, the Wannier centers for the first three out-of-plane bands degenerate at Berry phase $\phi = 0, \pi$, relating to a trivial second Stiefel-Whitney number, $w_2 = 0$. However, the induced coupling between the in-plane and out-of-plane modes in the mass-beam model disrupts such degeneracies, leading to odd crossing times at $\phi = \pi$ (see Supplementary Materials). This is equivalent to a nontrivial second Stiefel-Whitney number, $w_2 = 1$, defined as [21]

$$(-1)^{w_2} = \prod_{n=1}^{4}(-1)^{\lfloor N^-(\Lambda_n)/2 \rfloor} \quad (2)$$

where $N^-(\Lambda_n)$ represents the number of modes below the gap with negative parity at the time-reversal invariant momenta $\Lambda_n$, which are Γ and three M points in the Brillouin zone. The floor function $\lfloor \bullet \rfloor$ is rounded down to the nearest integer. The irreducible representations of the eigenmodes at the Γ point are 2D $E_+$ and 1D $A_{2-}$ representations, whereas at the M point, they are $A_-$, $B_+$, and $B_-$. It can be straightforwardly determined that $N^-(\Gamma) = 1$, and $N^-(M) = 2$ for each of the three M points. Then, the second Stiefel-Whitney number for the gap is $w_2 = 1$. As a result, a Stiefel-Whitney TI featuring a higher-order band topology appears in a vector-wave system.

Interestingly, a triple degeneracy at the Γ point, involving two linear in-plane Goldstone modes and one quadruple out-of-plane mode, is present at zero frequency. The node created by the two Goldstone modes has a quaternion frame charge $q = -1$ and a patch Euler class (see Supplementary Materials) $e_1 = \pm 1$, which is protected by the Nambu-Goldstone theorem and the absence of coupling between the linear Goldstone modes at the quadratic branch at Γ [42, 43]. Such unique triple degeneracy is also absent in scalar-wave systems. Meanwhile, another triple degeneracy occurs at the K point with a quaternion frame charge, $q = +k$, which plays a crucial role in the braiding, transfer, and merging of band nodes among the first three bands.

We have demonstrated Euler and Stiefel-Whitney TI phases in scalar and vectorial elastic systems, respectively. We now consider whether these two distinct topological classifications and phases can coexist within a single system. To explore this, we consider a system with stiffness parameters $\mathbf{D}_1 = [3.2]$, $\mathbf{D}_2 = [5.5]$, $\mathbf{D}_3 = [4.0]$, and $\mathbf{D}_4 = [7.0]$, where the bandgap between the first and second dispersions opens, indicating the emergence of the Euler TI phase (see Fig. 1(d)). Meanwhile, the Stiefel-Whitney topological bandgap, marked by the green region, remains present as well. While previous studies on scalar-wave fields, such as phononic crystals [27, 28], have realized Euler and Stiefel-Whitney TI phases individually, we have demonstrated, for the first time, the coexistence of both topological phases in a single material, thanks to the vectorial nature of the elastic material. To demonstrate the robustness of these two topological phases, we vary $\mathbf{D}_1$ and $\mathbf{D}_2$ while keeping $D_3 = [4.0]$ and $D_4 = [7.0]$ unchanged. The Euler, Stiefel-Whitney, and hybrid TI phase regions are depicted by the orange, green, and shaded areas in Figs. 1(e) and 1(f), respectively, demonstrating their topological protection against varying parameters. More importantly, when the parameters fall within the Stiefel-Whitney TI region but outside the hybrid TI region, the non-Abelian braiding behaviors of the first three bands in vector-wave systems can be observed (see Supplementary Materials).

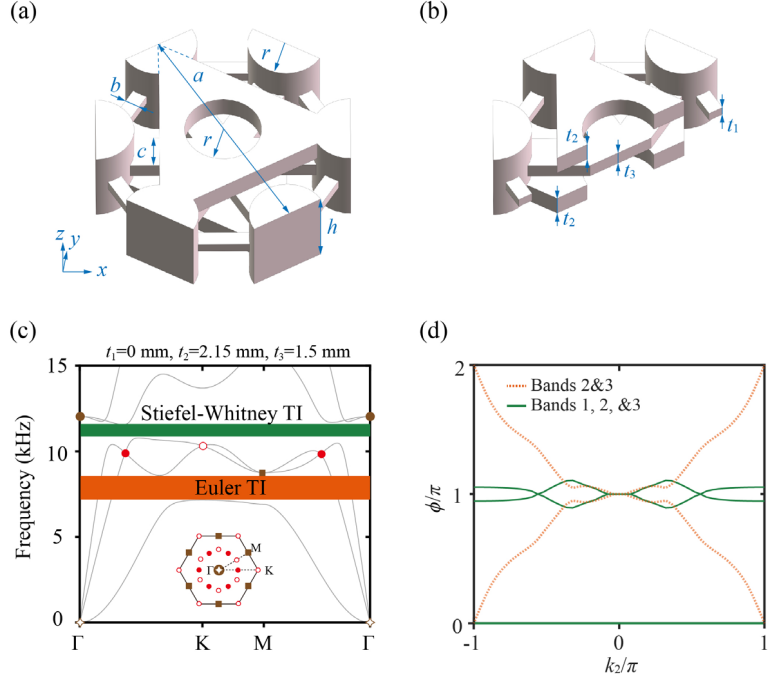

FIG. 2. Hybrid TI phases in elastic metamaterials. (a) and (b) depict the geometric parameters of the unit cell. (c), illustrate the evolution of dispersion curves and their corresponding nodes (insets) as the parameters are tuned. Different colors and fillings of the markers represent distinct frame charges and patched Euler classes. (d) The non-Abelian Wilson loops of different bands in (c).

*Hybrid TI phases in elastic metamaterials.* We now implement realistic material parameters that enable hybrid TI phases. The unit cell shown in Figs. 2(a) and 2(b) comprises of three cylinders and three links. The lattice constant is $a = 40/\sqrt{3}$ mm. The widths of the links are $b = 1.6$ mm and $c = 3.2$ mm for the NN and TNN connections, respectively. The cylinders have a height of $h = 8$ mm and a radius of $r = 3.5$ mm. Made of 316L stainless steel, the unit cell has a mass density of $\rho = 8000$ kg/m³, a Young's modulus of $E = 190$ Gpa, and a Poisson's ratio of $\mu = 0.26$.

The thicknesses of the links, as marked in Figs. 2(b), are denoted as $t_1$, $t_2$, and $t_3$ for the NN, NNN, and TNN connections, respectively, reflecting the coupling strength between the cylindrical masses. As these geometric parameters vary, the elastic metamaterial can experience topologically trivial, Stiefel-Whitney, and hybrid phases, as shown by the numerical results in the Supplementary Materials. Here, we focus on the geometric configuration used in the experimental fabrication, with $t_1 = 0$ mm (i.e., the NN links are disconnected with a vanishing coupling strength), $t_2 = 2.15$ mm, and $t_3 = 1.5$ mm. Mechanically, the TNN link also contributes to the NN coupling through indirect processes. The vectorial elastic band structure is numerically computed in Fig. 2(c), where the green and orange band gaps correspond to the Stiefel-Whitney and Euler TI phases, respectively.

The topological nature of these two distinct classes can be determined by calculating the non-Abelian Wilson loops of the band structures. In Fig. 2(d), the orange dashed lines represent a gapless Wilson loop for the second and third bands, where the nontrivial winding indicates an odd topological Euler class $e_1 = 1$. The green lines, corresponding to the Wilson loop for the first three bands, show three crossing points at $\phi = \pi$, signaling a topologically robust Stiefel-Whitney phase with $w_2 = 1$. This confirms that the hybrid TI phases, characterized by the Euler and Stiefel-Whitney classes, coexist within a single elastic metamaterial, as numerically demonstrated here. Very interestingly, the triply degenerate point at Γ has itself a topological stability captured by the nontrivial three-band quaternion charge −1. The decoupling of the linear Goldstone modes from the quadratic out-of-plane branch [43] readily implies that the two Goldstone modes must form a double nodal point at Γ with a finite patch Euler class of $e_1 = \pm 1$, i.e., it made of a superposition of two nodal points with identical non-Abelian frame charges. This further implies that the nontrivial global Euler topology of the two-band subspace (indicated by the Wilson loop winding over the whole Brillouin zone) but also the nontrivial Stiefel-Whitney topology of the three-band subspace (since it is effectively obtained from the two-band subspace by adding one single trivial band) are both directly originating from the topology of the Goldstone modes. Since the Goldstone mode crossing is actually protected by the fundamental Nambu-Goldstone theorem, we conclude that the nontrivial topology of the band structure is intrinsically unavoidable.

*Experimental demonstrations of hybrid TI phases.* Armed with the above theoretical analysis and numerical simulations, we fabricate an elastic metamaterial with hybrid Euler and Stiefel-Whitney TI phases using 3D metal-printing technology. A triangular sample with a side length of 20 unit cells is shown in Fig. 3(a). The cylinders on the boundary are half of the sizes. To measure the bulk dispersion curves, we use a piezoelectric ceramic exciter located at the center of the sample as a point-like source to generate elastic waves. To prevent interference from reflected waves at the boundaries, we apply ethylene-vinyl acetate copolymers to absorb excess elastic waves. We characterize the amplitude fields by measuring the sample's frequency responses through spatial scanning, with $a/2$ step along the *x*-axis and $\sqrt{3}a/2$ step along the *y*-axis. The amplitudes and phases are captured using a laser vibrometer combined with a network analyzer.

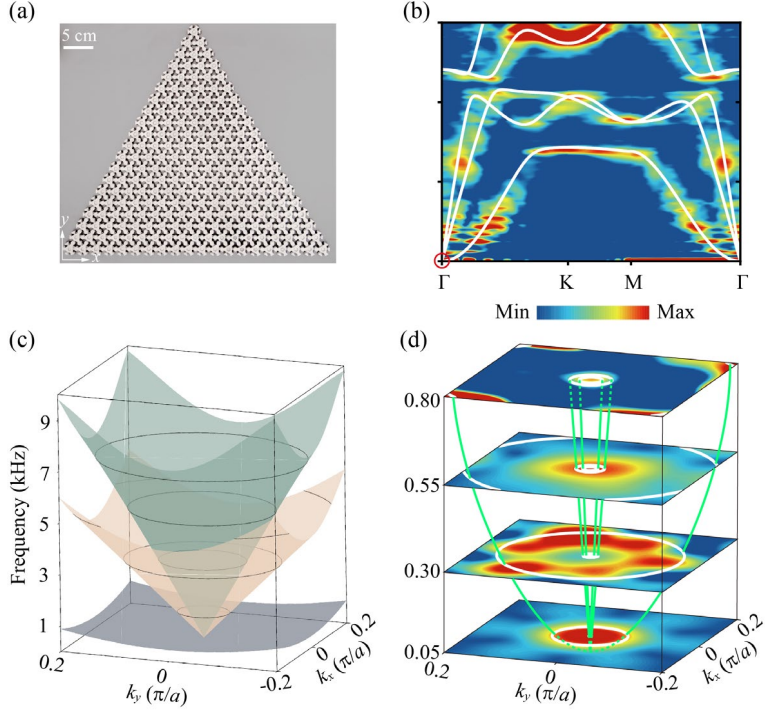

FIG. 3. Experimental protocol for the elastic metamaterial with hybrid TI phases. (a) Photograph of 3D metal-printed sample. (b) Measured band structures for full-vector modes of the elastic metamaterial. The solid lines represent the calculated dispersions and the color maps show the measured results. (c) and (d) show the measured and calculated linear dispersions around the Γ point, respectively. The white lines indicate the calculated iso-frequency contours, and the green lines serve as visual guides to highlight the calculated dispersion.

To fully observe the hybrid TI phases, we tilt the laser vibrometer at an 30° angle relative to the plane of the sample. This adjustment allows the measured displacements and phases to capture both in-plane and out-of-plane motions. The full-vector dispersion curves are obtained using 2D Fourier transformations, as shown in Fig. 3(b). In addition to the out-of-plane modes, the in-plane modes are also clearly identified. Two distinct gaps, corresponding to the Euler and Stiefel-Whitney TIs, are clearly visible. The excellent agreement between the calculated bands and measured dispersion curves demonstrates the high precision of the fabricated sample.

The triple degenerate node at the Γ point, with a frame charge $q = -1$ and Euler class $e_1 = \pm 1$, is depicted in Fig. 3(c) (calculated) and 3(d) (measured). The linear Dirac node at the K point characterized by a frame charge $q = -j$, and the quadratic node at the M point with a frame charge $q = -1$ and Euler class $e_1 = 1$ are presented in the Supplementary Materials. These unique dispersions and their associated non-Abelian frame charges reveal the abundant properties of the elastic metamaterial with hybrid TI phases.

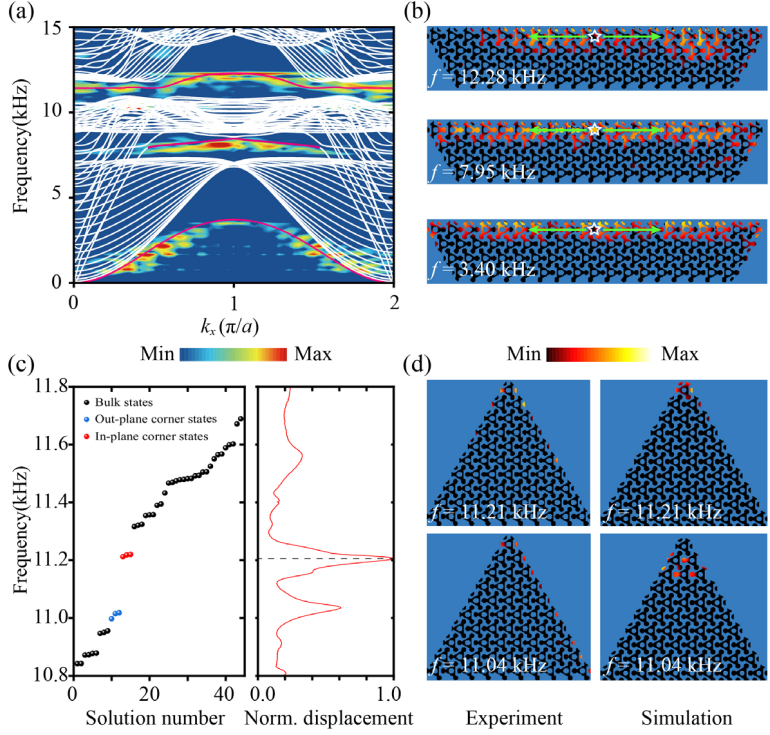

Fig. 4. Measured edge and corner states in elastic metamaterials with hybrid phases. (a) Dispersion projected onto the zigzag boundary, with the measured results shown as a color map and the simulated data as dots. The red curves indicate edge states. (b) Measured displacement distribution along the free boundary with a white star indicating a point-like source. (c) Left panel: Phononic eigenfrequency spectrum of triangular structure. Right panel: source-to-detector response spectra for corner measurements. (d) Measured and simulated displacement profiles of corner states.

The emergence of edge states indicates the Euler TI phase, while the presence of edge states and in-gap corner states is characteristic of the Stiefel-Whitney TI phase. To confirm these two phases, we measure the topological edge and corner states in experiments. Fig. 4(a) shows the projected dispersion of edge states along the zigzag boundary, with experimental data depicted in color maps and white dots representing the numerical results, demonstrating excellent agreement. In Fig. 4(b), the spatial profiles of mode displacements at the in-gap frequency, 7.99 kHz and 12.28 kHz, reveal bidirectional wave transmission due to the time-reversal symmetry. Further details on edge state propagation along the armchair boundary are provided in the Supplementary Materials. The coexistence of these two distinct types of in-gap edge states offers strong evidence for the hybrid Euler and Stiefel-Whitney TI phase with $e_1 = 1$ and $w_2 = 1$. Additionally, as shown in Figs. 4 (a) and 4(b), edge states are observed at frequencies below the bulk band, corresponding to Rayleigh surface waves at the boundary [42, 44].

Fig. 4(c) shows numerical calculations of the elastic band spectra for a triangular sample, revealing three in-plane (red) and three out-of-plane (blue) corner states within the bandgap. By applying a point shaking source three unit cells away from the corner, these topological states can be excited. Two resonance peaks at 11.21 kHz and 11.04 kHz are observed in the source-to-detector curve (right panel of Fig. 4(c)), corresponding to the corner states shown in Fig. 4(d). The excellent agreement between experimental and simulation results confirms the emergence of corner states linked to the Stiefel-Whitney TI phase. Furthermore, by introducing defects near the corners, the existence and frequency the corner states cannot be easily modified, which demonstrates the robustness of the corner states (see Supplementary Materials).

*Conclusion.* In summary, we have theoretically formulated a mass-beam model to predict the existence of hybrid Euler and Stiefel-Whitney TI phases by leveraging the coupling between in-plane and out-of-plane modes, and then experimentally realized an elastic metamaterial with hybrid TI phases in a metal-printed sample. We revealed the key role played by the nontrivial topological nature of the Goldstone modes' crossing at Γ, equivalently characterized by a three-band quaternion charge −1 and a finite two-band patch Euler class. Through consistent theory, simulations, and experiments, we confirm the existence of gapped 1D edge states and 0D higher-order corner states due to an odd Euler class and a nontrivial second Stiefel-Whitney number. Experimental discovery using elastic metamaterials provides a valuable platform that could inspire further exploration of novel topological classes, leading to new possibilities and applications such as bifunctional devices for sensing and wave manipulation in solid mechanics, topological physics, and materials science.

**Acknowledgments**

Y. W. would like to thank Prof. Jian-Hua Jiang, Prof. Zhen Gao, and Dr. Zhi-Kang Lin for helpful discussions. The authors wish to acknowledge the support from the National Nature Science Foundation of China (Grant Nos. 12272040, 12374157, and 12302112), and the Fundamental Research Funds for the Central Universities (No. 30923010207). Adrien Bouhon was partially funded by a Marie-Curie fellowship (101025315) and acknowledged financial support from the Swedish Research Council (Vetenskapsradet) (2021-04681). Robert-Jan Slager acknowledged funding from an EPSRC ERC underwrite grant EP/X025829/1, and a Royal Society exchange grant IES/R1/221060, as well as Trinity College, Cambridge.